\documentclass[%
 reprint,
superscriptaddress,
 amsmath,amssymb,
 aps,
pra,
]{revtex4-2} 
\DeclareUnicodeCharacter{02BC}{'}

\usepackage{ulem}
\usepackage{comment}
\usepackage{graphicx}
\usepackage{booktabs}
\usepackage{dcolumn}
\usepackage{bm}
\usepackage[dvipsnames]{xcolor}
\usepackage{newtxmath}
\usepackage{mathtools}
\usepackage{physics}
\usepackage{float}
\usepackage[caption = false]{subfig}
\usepackage{tabularx}
\usepackage{pifont}

\newcommand{\be}{\begin{equation}}
\newcommand{\ee}{\end{equation}}

\def\qt{{\tilde{q}}}

\begin{document}

\preprint{APS/123-QED}

\title{Anisotropic scale invariance and the uniaxial Lifshitz point from the nonperturbative renormalization group}

\author{Gonzalo De Polsi } 
\email{gonzalo.depolsi@fcien.edu.uy}
\affiliation{Instituto de Física, Facultad de Ciencias, Universidad de la República,\\ Iguá 4225, Montevideo, Uruguay}
\author{Pawel Jakubczyk} 
\email{pawel.jakubczyk@fuw.edu.pl}
\affiliation{Institute  of Theoretical Physics, Faculty of Physics, University of Warsaw, Pasteura 5, 02-093 Warsaw, Poland}

\date{\today}

\begin{abstract}
We employ the derivative expansion of the nonperturbative renormalization group
to address the phenomenon of anisotropic scale invariance and the associated functional fixed points, also known as Lifshitz points, in systems characterized by a scalar order parameter.  
We demonstrate the existence of the Lifshitz fixed point featuring a non-classical value of the anisotropy exponent $\theta<1/2$ and provide  estimates for values of a set of critical exponents in the physically most relevant case of the three-dimensional uniaxial Lifshitz point  $(d,m)=(3,1)$, $m$ denoting the anisotropy index. We compare our predictions with existing estimates from perturbative expansions around dimensionality $d=4+\frac{1}{2}$ as well as those from the $1/N$ expansion. 
\end{abstract}

\maketitle

\section{\label{sec:INTRO} Introduction}
The phenomenon of scale invariance is amply represented and broadly studied in a diversity of systems of prior relevance to condensed matter and high-energy physics \cite{Goldenfeld_1992, Cardy_1996, Zinn-Justin_2010}. The somewhat less explored and theoretically less understood case concerns anisotropic scale invariance \cite{Chaikin_1995, Diehl_2002, Pisarski_2019}, where spatial anisotropy of a microscopic system is not smoothened by coarse-graining \cite{Chen_2004}, and achieving scale invariance requires applying the scaling transformation to different directions of the physical space in distinct ways. Such a situation has since long been known to arise in a variety of physical systems including {\it inter alia} magnets \cite{Becerra_1980, Butera_2008, Plackowski_2011}, liquid crystals \cite{Chen_1976, Chaikin_1995}, polymer systems \cite{Dobrynin_1996, Schwahn_1999, Pisarski_2019}, membranes \cite{Radzihovsky_1995, Essafi_2011} or quantum chromodynamics \cite{Casalbuoni_2004}. More recently, such setups were also invoked in the contexts of general quantum criticality \cite{Ramazashvili_1999, Gunther_2007,  Zhao_2020}, frustrated magnetism \cite{Balents_2016, Kharkov_2020}, altermagnetism \cite{Hu_2025}, unconventional superconductivity \cite{Waardh_2018} and superfluidity of ultra-cold atomic Fermi \cite{Gubbels_2009, Pimenov_2018, Zdybel_2020, Pini_2021, Zdybel_2021} as well as Bose \cite{Jakubczyk_2018, Lebek_2020, Lebek_2021} gases. For equilibrium condensed matter systems anisotropic scale invariance  occurs \cite{Chaikin_1995, Diehl_2002}, in particular, at the Lifshitz points, i.e. critical points involving three thermodynamic phases, one of them being disordered, one uniformly ordered and the remaining one non-uniformly ordered (i.e. exhibiting modulation of the order parameter). In the standard magnetic setup this corresponds to points in the phase diagram, where paramagnetic, ferromagnetic and antiferromagnetic phases meet. This situation occurs, as already indicated, in ample quantum matter systems being of major present interest. On the other hand, the concept of the Lifshitz point is not at all new and was first discussed already in 1970s \cite{Hornreich_1975, Michelson_1977, Michelson_1977_2, Grest_1978} in relation primarily to magnets. Its most well-studied microscopic realization is presumably in the anisotropic next-nearest-neighbor Ising (ANNNI) model \cite{Selke_1988}. Despite this long history and the wealth of distinct realizations, description of the Lifshitz points via the renormalization group (RG) tools may be characterized as  largely underdeveloped as compared to the isotropic systems. This situation is not hard to explain considering the amount of technical difficulties involved. Notably, the uniaxial Lifshitz point has been perturbatively studied only up to two-loop order \cite{Mergulhao_1999, Diehl_2000, Shpot_2001} and the upper critical dimension in question is $d_u=4+\frac{1}2$, such that extrapolation of the expansion in $\epsilon =d_u-d$ to $d=3$ is more problematic as compared to the standard isotropic cases, where $d_u^{(isot)}=4$. Another notable fact is that the Lifshitz exponents have been computed within the $1/N$ expansion only up to terms of order $1/N$ \cite{Shpot_2005, Shpot_2008} and these two abovementioned approaches, when extrapolated to $(d,N)=(3,1)$, give conflicting results concerning even the sign of one of the exponents. 
Relatively recently, the anisotropic Lifshitz point was also  addressed in the framework of nonperturbative RG \cite{Essafi_2012, Zdybel_2021}. The latter of these two contributions concentrated exclusively on the case $N=3$ and used a rather simple and unsystematic approximation, while the former one was restricted to the leading order of the derivative expansion (DE) [the so-called local potential approximation (LPA) or $\partial^0$] and its minimal extension (referred to as "constrained LPA' "), which mistreated one of the relevant RG eigenperturbations. The purpose of the present work is to approach the anisotropic Lifshitz point within the DE retaining all the symmetry-allowed terms at order $\partial^2$ as well as the physically most relevant terms of order $\partial^4$, which provides a more systematic and controlled framework. We stress that, in contrast to the standard isotropic situations, inclusion of the $\partial^4$ term is mandatory due to the physical nature of the problem. 

The outline of the present paper is as follows: in Sec.~\ref{sec:GLmodel} we introduce and review the analyzed Landau-Ginzburg type model, in Sec.~\ref{sec:NPRG} we discuss an  adaptation of the derivative expansion of the nonperturbative RG to anisotropic situations and explain how our equations lead to anisotropic scaling solutions. In Sec.~\ref{sec:DE2pW} we analyze the structure of the flow equations at LPA level, leading to  analytical insights and approximate relations with the isotropic setups in reduced dimensionalities, followed by the presentation of our main results concerning the scaling solutions and critical exponents retaining the functional RG flow of the gradient and laplacian terms in the effective action. We summarize the paper in Sec.~\ref{sec:concl}.

\section{The Landau-Ginzburg model and anisotropic scale invariance} \label{sec:GLmodel}
We discuss the anisotropic scale invariance in the framework of the Landau-Ginzburg model defined by the bare action \cite{Chaikin_1995, Diehl_2002, Pisarski_2019}: 
\begin{equation}
S[\phi]=  \int d^d x \left[U^0(\phi^2)+\frac{1}{2}(\nabla_{\perp}\phi)^2 + \frac{1}{2}Z_{\parallel}^0(\nabla_{\parallel}\phi)^2 + \frac{1}{2}W_{{\parallel}}^0 (\Delta_{\parallel}\phi)^2  \right]\;, 
\label{LGW}
\end{equation}
where $\phi$ is a real scalar order parameter, and the spatial coordinates involve two classes of components such that  
\be
d^dx=d^mx_{\parallel}d^{d-m}x_{\perp}\;,
\ee
\be
(\nabla_{\parallel}\phi)^2=\sum_{\alpha=1}^{m}\left(\frac{\partial \phi}{\partial x_\alpha}\right)^2\;, \;\;\;\; 
(\nabla_{\perp}\phi)^2=\sum_{\alpha=m+1}^{d}\left(\frac{\partial \phi}{\partial x_\alpha}\right)^2\;,
\ee
and $m\in\{0,1,\dots d\}$ is the anisotropy index. In the magnetic language, the $m$ distinct space directions represent the ones, to which the possible modulation of the order parameter is constrained. Our major focus is the uniaxial case $m=1$, however, at formal level, we allow $m$ as well as $d$ to vary continuously. The effective potential $U^0(\phi^2)$ takes the standard form of a quartic polynomial in $\phi$:
\be
U^0(\phi^2) = \tau \phi^2 + u \phi^4 
\ee
with $u>0$. We assume $W_{{\parallel}}^0 > 0$ and the field $\phi$ has been defined such that the coefficient of $(\nabla_{\perp}\phi)^2$ in Eq.~(\ref{LGW}) is 1/2.  

For $Z_{\parallel}^0>0$, as far as universal bulk properties are concerned,  there is no essential difference between the two classes of spatial directions and, at mean-field (MF) level,  the Laplacian term $W_{{\parallel}}^0 (\Delta_{\parallel}\phi)^2$ can be completely disregarded. By varying $\tau$ the system is then tuned through a phase transition of ferromagnetic type and, in the RG description, the critical singularities are controlled by the standard Wilson-Fisher fixed point. For $Z_{\parallel}^0<0$ it is energetically beneficial for the system to create a spatial modulation of the ordering field $\phi$ and the magnitude of the ordering wavevector follows from the balance between the gradient and laplacian terms in the $\parallel$ direction. By varying $\tau$ one then tunes the system through a transition between the paramagnetic and antiferromagnetic phases, which is again controlled by the Wilson-Fisher RG fixed point. A special situation occurs (at MF level) for $Z_{\parallel}^0=0$, where, for $\tau=0$, the system exhibits an anisotropic scale-invariant critical point (the Lifshitz point) \cite{Chaikin_1995, Diehl_2002, Pisarski_2019}. In particular the MF correlation function 
\be 
\label{GMF}
\langle\phi_{\vec{q}} \phi_{-\vec{q}} \rangle =G(\vec{q})=G(\vec{q}_\perp, \vec{q}_\parallel) =\frac{1}{Z_\perp \vec{q}_\perp^2 + W_{{\parallel}}^0 (\vec{q}_\parallel^2)^2}
\ee
depends on the direction of momentum $\vec{q}$. In the vicinity of the Lifshitz point there are two correlation length exponents governing the singularity of the correlation lengths, such that 
\be 
\xi_\perp \sim |\tau|^{-\nu_{\perp}}\;, \;\;\textrm{and} \;\;\xi_\parallel \sim |\tau|^{-\nu_{\parallel}}\;.
\ee 
Once again, at MF level one finds $\nu_\parallel=\frac{1}{2}\nu_\perp =\frac{1}{4}$. Generically, it can be shown that the two exponents obey the scaling relation \cite{Diehl_2002}
\be 
\label{scaling}
\nu_\parallel=\frac{2-\eta_\perp}{4-\eta_\parallel}\nu_\perp = \theta \nu_\perp\;,
\ee
which defines the anisotropy exponent $\theta$. The quantities $\eta_\perp$ and $\eta_{\parallel}$ are the anomalous dimensions corresponding to the distinct inequivalent spatial directions. At Lifshitz criticality one anticipates 
\be 
G^{-1}(\vec{q}_\perp, 0)\sim |\vec{q}_\perp|^{2-\eta_\perp}
\ee 
and
\be 
G^{-1}(0, \vec{q}_\parallel)\sim |\vec{q}_\parallel|^{4-\eta_\parallel}
\ee
for small momenta.

\section{Nonperturbative RG and the derivative expansion} \label{sec:NPRG}

As emphasized, in spite of the ample realizations of the Lifshitz criticality in condensed-matter systems, the level of development of the RG (or any other) theory of these phenomena, both in the perturbative and non-perturbative versions, is way lower than for the isotropic cases. We now discuss how the nonperturbative RG calculation is carried out.

Our present approach to the Lifshitz point departs from the Wetterich equation \cite{Wetterich_1993}
\be 
\label{Wetterich}
\partial_k \Gamma_k [\phi]=\frac{1}{2}{\rm Tr}\left[ \partial_k R_k(\vec{q}) \left( \Gamma_k^{(2)}[\phi]+R_k(\vec{q})\right)^{-1} \right]\;.
\ee
Here $\Gamma_k [\phi]$ denotes the effective action in presence of a (momentum) cutoff at scale $k$, $\Gamma_k^{(2)}[\phi]$ is its second (functional) derivative with respect to $\phi$, and $R_k(\vec{q})$ represents the cutoff function suppressing propagation of modes with momentum $\vec{q}$ above the cutoff scale $k$. For $k\to k_{UV}$ the quantity $\Gamma_k [\phi]$ approaches the bare action $S[\phi]$ such as the one given in Eq.~(\ref{LGW}), while for $k\to 0$ (cutoff removed) we find $\Gamma_k [\phi] \to F[\phi]$, where $F[\phi]$ represents the Gibbs free energy of the statistical-mechanical system defined by $S[\phi]$. The derivative expansion is among the most broadly implemented and successful approximations in this framework. The order $\partial^n$ of the DE consists in projecting the flowing functional $\Gamma_k[\phi]$ onto a set of scale-dependent functions retaining only symmetry-allowed terms involving at most $n$ derivatives of $\phi$. For example, for the Lifshitz point at order $n=4$ we have:  
\begin{widetext}
\begin{equation}
\begin{split}
\Gamma_k^{DE4}[\phi]=&\int d^dx \Bigg\{ U(\rho)+\frac{1}{2}Z_{\perp}(\rho) (\nabla_\perp \phi)^2 + \frac{1}{2}Z_{\parallel}(\rho) (\nabla_\parallel \phi)^2 +\frac{1}{2} W_{\perp} (\rho) (\Delta_{\perp}\phi)^2+\frac{1}{2} W_{\parallel} (\rho) (\Delta_{\parallel}\phi)^2 + \\ &\frac{1}{2} X_{\perp} (\rho)\phi (\nabla_{\perp}\phi)^2(\Delta_\perp \phi) + 
\frac{1}{2} X_{\parallel} (\rho)\phi (\nabla_{\parallel}\phi)^2(\Delta_\parallel \phi) + 
\frac{1}{2} Y_{\perp} (\rho)[(\nabla_{\perp}\phi)^2]^2 + \frac{1}{2} Y_{\parallel} (\rho)[(\nabla_{\parallel}\phi)^2]^2\\ &+\frac{1}{2} W_{\angle} (\rho)(\Delta_\parallel \phi)(\Delta_\perp \phi) + 
\frac{1}{2} X_{1,\angle} (\rho)\phi (\nabla_{\parallel}\phi)^2(\Delta_\perp \phi)+ 
\frac{1}{2} X_{2,\angle} (\rho)\phi (\nabla_{\perp}\phi)^2(\Delta_\parallel \phi) + 
\frac{1}{2} Y_{\angle} (\rho)(\nabla_{\parallel}\phi)^2(\nabla_{\perp}\phi)^2
\Bigg\}\;.
\end{split}
\label{DE4}
\end{equation}
\end{widetext}
We suppressed the $k$-dependence of the parametrizing functions for clarity of notation and introduced $\rho=\phi^2/2$. Note that the Ansatz of Eq.~(\ref{DE4}) involves no field truncation and encompasses both the isotropic and anisotropic situations. By inserting the derivative expansion Ansatz into the Wetterich equation [Eq.~(\ref{Wetterich})] one obtains the flow of the parametrizing functions. For the standard $O(N)$ models \cite{Balog_2019, Polsi_2020, Polsi_2021, Chlebicki_2022}, the $q$-state Potts models \cite{Villalobos_2023}, or frustrated magnets featuring the $O(N)\times O(2)$ symmetry \cite{Villalobos_2025}, systematic implementation of the DE leads to very accurate results concerning universal critical properties.  
  
In the practical calculations to follow later in this paper, we will drop the functions $X_\perp$, $X_\parallel$, $X_{1,\angle}$, $X_{2,\angle}$, $Y_\perp$, $Y_\parallel$, $Y_\angle$, $W_\angle$ and $W_\perp$ in Eq.~(\ref{DE4}), but will retain all the remaining field and scale dependencies. We refer to this approximation as ``DE(2)+W$_\parallel (\rho)$ truncation". The fluctuations in the $\perp$ direction are controlled by the gradient term involving $Z_\perp$ and the neglected terms of order $q_\perp^4$ are then small. The functions $X_{\perp/\parallel/i,\angle}$ and $Y_{\perp/\parallel/\angle}$ are involved in the 3- and 4- point functions [$\Gamma^{(3)}$ and $\Gamma^{(4)}$], respectively, and do not occur in the propagator of the full derivative expansion Ansatz (at any order). The term involving $W_\parallel (\rho)$ must be retained for obvious physical reasons (see Sec.~\ref{sec:GLmodel}). 
Obtaining the Lifshitz point via the RG flow from a given microscopic action requires tuning two parameters of the bare action, such that the order parameter mass, as well as the gradient coefficient $Z_\parallel$ both vanish in the limit $k\to 0$. In consequence, we anticipate two relevant RG directions at the Lifshitz fixed point. Although we will concentrate on the flow of the Ansatz functions as the scale $k$ of the regulator is varied, we stress  that the practical implementation will not consist in fine-tuning two parameters (which is technically quite problematic), but instead will concentrate on finding the fixed point solutions of the dimensionless flow equations starting from a suitable initial guess.

We now focus on the flow of the effective potential $U(\rho)$, which is obtained by first computing $\Gamma^{(2)}$ from the derivative expansion Ansatz, then plugging the result into Eq.~(\ref{Wetterich}) and evaluating at a spatially uniform field configuration. This leads to 
\be 
\label{Uflow}
\partial_k U(\rho)=\frac{1}{2} \int \frac{d^dq}{(2\pi)^d}\partial_k R_k(\vec{q})G_0(\vec{q},\rho)
\ee 
with 
\begin{eqnarray} 
\label{G0}
G_0^{-1}(\vec{q},\rho) &=&  \\
U'(\rho)&+&2\rho U''(\rho)+Z_\perp(\rho)\vec{q}_\perp^2+Z_\parallel(\rho)\vec{q}_\parallel^2+W_\parallel(\rho)\vec{q}_\parallel^4+R_k(\vec{q})\;.  \nonumber  
\end{eqnarray}
Eq.~(\ref{Uflow}) holds at any order of the DE up to adding [in Eq.(\ref{G0})] the terms with $W_\angle$, $W_\perp$ (which we disregard here) and terms involving higher powers of momenta in $G_0$. In Eq.~(\ref{G0}) we give the expression pertinent to the DE(2)+W$_\parallel (\rho)$ truncation studied in Sec.~\ref{sec:DE2pW}, i.e. the neglected terms are of order $\vec{q}_\parallel^2\vec{q}_\perp^2$, $\vec{q}_\perp^4$ and $\vec{q}_\parallel^6$. 

In order to seek for RG fixed point solutions, we perform the following rescaling: 
\begin{eqnarray} 
\vec{q}_\perp &=& k \tilde{\vec{q}}_\perp\;,\;\;\; \vec{q}_\parallel =k^\theta \tilde{\vec{q}}_\parallel\;,\;\;\; \rho =Z^{-1}k^{d-2+m(\theta-1)}\tilde{\rho}\;, \nonumber \\ 
U&=&k^{d+m(\theta-1)}\tilde{u}\;,\;\;\; Z_\perp=Z\tilde{z}_\perp\;,\;\;\; Z_\parallel=Z k^{2(1-\theta)}\tilde{z}_\parallel \nonumber \\ 
W_\parallel&=&Z k^{2(1-2\theta)}\tilde{w}\;,\;\;\; R=Zk^2 r\;, 
\label{rescaling}
\end{eqnarray}
which absorbs all the explicit dependencies on $k$ in the flow equations for the parametrizing functions. Note that the anisotropy exponent $\theta$ is unspecified and actually represents one of the key quantities we set out to compute; while $Z$ is the (scale-dependent) field rescaling factor.  

In terms of the rescaled variables, Eq.~(\ref{Uflow}) becomes: 
\begin{widetext}
\begin{eqnarray} 
\label{uflow}
\partial_t \tilde{u} = -\left(d+m(\theta-1)\right)\tilde{u}+\left(\eta_\perp+d+m(\theta-1)-2\right)\tilde{\rho}\tilde{u}'+ 
\mathcal{V}_{d,m}
\int_0^\infty d\qt_{\perp}\int_0^\infty d\qt_{\parallel} \qt_{\perp}^{d-m-1}\qt_{\parallel}^{m-1}\frac{(2-\eta_\perp)r-2\qt^2\partial_{\qt_{\perp}^2}r-2\theta\qt_\parallel^2\partial_{\qt_{\parallel}^2}r}{\tilde{z}_\perp\qt_\perp^2+\tilde{z}_\parallel\qt_\parallel^2+\tilde{w}\qt_\parallel^4+\tilde{u}'+2\tilde{\rho}\tilde{u}''+r}\;, \nonumber   \\ 
\end{eqnarray}    
\end{widetext} 
where $t=\ln(k/k_{UV})$, $\eta_\perp=-k\partial_t\ln Z$ is the running anomalous dimension corresponding to the $\perp$ direction, and $\mathcal{V}_{d,m}=\frac{1}{2}\frac{\mathcal{S}^{m-1}\mathcal{S}^{d-m-1}}{(2\pi)^d}$ absorbs constants emerging from doing the angular integrations in the two momentum subspaces (here $\mathcal{S}^{D}$ is the suface area of a unit $D$-dimensional sphere). Eq.~(\ref{uflow}) straightforwardly generalizes to higher order DE truncations simply by adding terms involving higher powers of momentum (in the $\perp$ and $\parallel$ directions) in the denominator on its right-hand side. It is evident that Eq.~(\ref{uflow}) is not closed and must be supplemented by the flow equations for $\tilde{z}_\perp$, $\tilde{z}_\parallel$, and $\tilde{w}$ (which are functions of both $k$ and $\tilde{\rho}$) - see Sec.~\ref{sec:DE2pW} and supplemental material. 

We additionally observe, that Eq.~(\ref{uflow}) contains the standard isotropic case at level DE(2), which is recovered by taking $m=0$ or $\theta=1$ and setting $\tilde{w}=0$.  

A substantial simplification due to restoration of isotropy occurs also for $m=d$ (the so-called isotropic Lifshitz point). This case is of interest for relativistic theories in $d\geq 4$ and was addressed with nonperturbative RG in Refs.~ \cite{Zappala_2017, Zappala_2018, Defenu_2021}. We will not discuss this situation in the present paper.

We now proceed to Sec.~\ref{sec:DE2pW}, where we summarize the LPA approximation, which closes Eq.~(\ref{uflow}), and subsequently we analyze the DE(2)+W$_\parallel(\rho)$ truncation.

\section{Lifshitz point in the  DE(2)+W$_{\parallel}(\rho)$ truncation}\label{sec:DE2pW} 

In this section we present the results stemming from the nonperturbative RG and compare to available results in the literature. To this end, we first discuss the LPA approximation which was previously implemented in Ref.~\cite{Zdybel_2021} and we highlight the differences and the generalization used for the more reliable DE(2)+W$_{\parallel}(\rho)$ approximation. Subsequently, we present the results obtained with nonperturbative RG for the anisotropic fixed point with $m=1$ and we compare to the literature. Lastly, we discuss the error bar estimation procedure.

\subsection{The local potential approximation and moving to higher order} \label{ssec:LPA}

We start by reviewing the LPA ($\partial^0$) truncation, where the flow of momentum dependencies of $\Gamma^{(2)}$ is disregarded. In particular, we impose $\tilde{z}_\parallel=0$. This closes Eq.~(\ref{uflow}), which can then be studied numerically.

The LPA approximation neglects the anomalous dimensions and implies $\theta=1/2$. We invoke the observation of  Ref.~\cite{Zdybel_2021} that by taking the dimensionless cutoff $r$ [compare Eq.~(\ref{rescaling})] in the following form: 
\be 
\label{cutoffLPA}
r(\vec{\qt}_{\perp}, \vec{\qt}_{\parallel}) = r(\vec{\qt}_{\perp}^2+\vec{\qt}_{\parallel}^4)
\ee 
and performing a suitable rescaling of $\tilde{\rho}$ and $\tilde{u}$, one maps Eq.~(\ref{uflow}) to its isotropic counterpart in dimensionality $d_{{\rm eff}}=d-\frac{m}{2}$. This result implies that the LPA-level fixed point as well as the critical exponents of the $m$-axial Lifshitz point in $d$ dimensions coincide with those of the $O(N)$ universality class in dimensionality decreased by $m/2$. Even though this correspondence does not hold beyond LPA for $d-\frac{m}{2}<4$, it becomes increasingly accurate when approaching the upper critical dimension, where anomalous dimensions vanish. We note that the isotropic $O(1)$ fixed point features only one relevant direction in contrast to the Lifshitz point, which has two; as a consequence, the LPA-level correspondence must, in fact, be violated in an exact treatment.

The above insights, borrowed from Ref.~\cite{Zdybel_2021} inspire our strategy of searching for the anisotropic fixed points at order DE(2)+W$_{\parallel}(\rho)$ which is adopted in Sec.~\ref{sec:DE2pW}. This relies on the above invoked correspondence and exploits its increasing  accuracy upon elevating the system dimensionality towards $d_u=4+\frac{m}{2}$.

When elevating the truncation to order DE(2)+W$_{\parallel}(\rho)$, in order to compute the anisotropic scaling solutions the flow of the effective potential [Eq.~(\ref{uflow})] must be supplemented with the flow equations for the parametrizing functions $\tilde{z}_\perp(\tilde{\rho})$, $\tilde{z}_\parallel(\tilde{\rho})$, and $\tilde{w}(\tilde{\rho})$. These are derived  by the procedure well described e.g. in Ref.~\cite{Dupuis_2021} for the isotropic situations. By taking consecutive derivatives of the DE(2)+W$_{\parallel}(\rho)$ Ansatz and evaluating at uniform field configuration, we obtain the vertex functions $\Gamma^{(2)}$, $\Gamma^{(3)}$, $\Gamma^{(4)}$, which depend on  $\rho$ and momenta. Subsequently,  by differentiating Eq.~(\ref{Wetterich}) twice with respect to $\phi$ we obtain the flow equation for the 2-point function $\Gamma^{(2)}(p_\perp^2,p_\parallel^2, \rho)$, which involves $\Gamma^{(3)}$ and $\Gamma^{(4)}$.  In the  next step we expand in $p_\perp^2$ and $p_\parallel^2$ and identify the coefficients of the expansion proportional to $p_\perp^2$, $p_\parallel^2$, and $p_\parallel^4$ as flow equations for $Z_\perp(\rho)$, $Z_\parallel(\rho)$, and $W_\parallel(\rho)$, respectively. After performing the rescaling dictated by Eq.~(\ref{rescaling}), we obtain the flow equations characterized by the following structure: 
\begin{eqnarray} 
\label{floweq}
\partial_t\tilde{z}_\perp&=&\eta_\perp \tilde{z}_\perp + \left(\eta_\perp+d-2+m(\theta-1)\right)\tilde{\rho}\tilde{z}_\perp'+{\rm L}_{\tilde{z}_\perp}  \\
\partial_t\tilde{z}_\parallel&=&\left(\eta_\perp-2(1-\theta)\right)\tilde{z}_\parallel+\left(\eta_\perp+d-2+m(\theta-1)\right)\tilde{\rho}\tilde{z}_\parallel'+{\rm L}_{\tilde{z}_\parallel} \nonumber \\ 
\partial_t\tilde{w}&=&\left(\eta_\perp-2(1-2\theta)\right)\tilde{w}+\left(\eta_\perp+d-2+m(\theta-1)\right)\tilde{\rho}\tilde{w}'+{\rm L}_{\tilde{w}_\parallel} \nonumber\;.
\end{eqnarray}
The lengthy loop contributions ${\rm L}_{\tilde{z}_\perp}$, ${\rm L}_{\tilde{z}_\parallel}$, and ${\rm L}_{\tilde{w}_\parallel}$  involve two-dimensional momentum integrals and are explicitly given in the supplemental material. From a technical point of view, the necessity of dealing with controlled numerical two-dimensional momentum integrations constitutes the major challenge as compared to the isotropic situations. Note that the previous studies evaded this complication either by regularizing only some of the  directions in momentum space \cite{Essafi_2012}, or by disregarding the flow of $Z_\parallel (\rho)$ \cite{Zdybel_2021}.
The exponents $\eta_\perp$ and $\theta$ are determined from the normalization condition which fixes the values of $\tilde{z}_\perp(\tilde{\rho}_0)=1$ and $\tilde{w}(\tilde{\rho}_0)=1$ at an arbitrary point $\tilde{\rho}_0$.  

By putting $\tilde{w}=0$ and imposing $\theta=1$ we restore symmetry between the $\perp$ and $\parallel$ directions and the flow equations then reduce to the isotropic DE(2) case. We, however, presently seek for a completely different class of fixed point solutions of Eq.~(\ref{floweq}), which, at least for $d\approx d_u=4+m/2$, exhibit $\theta\approx 1/2$. 

The cutoff characterized by Eq.~(\ref{cutoffLPA}) is not necessarily the most suitable for the present general case, where $\theta\neq 1/2$. In what follows we take 
\be
\label{cutoffDE}
r(\vec{\qt}_{\perp}, \vec{\qt}_{\parallel}) = \alpha r_0(\vec{\qt}_{\perp}^2)r_0(\vec{\qt}_{\parallel}^4) 
\ee
with $\alpha$ representing a variable parameter (see below). 
As concerns the form of the function $r_0$, we investigated three choices: the Wetterich regulator \cite{Berges_2002}:
\be 
r_0(x)= r_{W}(x)=\frac{x}{e^x-1}\;, 
\ee 
the exponential regulator 
\be
r_0(x)=r_{E}(x)=e^{-x}\;,
\ee
and the ''$\theta_3$'' regulator 
\be 
r_0(x)=r_{\theta_3}(x)=(1-x)^3\theta (1-x)\;.
\ee
The status of the exponential regulator $r_{E}(x)$ is distinct in the present context since it automatically obeys Eq.~(\ref{cutoffLPA}). The ``$\theta_3$'' regulator is a smoothened version of the Litim regulator \cite{Litim_2001}, suitable for analysis reaching above the DE(2) truncation level, for a discussion, see \cite{Polsi_2020}.
As already mentioned, our cutoff is  entirely different from the one used in Ref. \cite{Essafi_2012}, which regularized only the $\parallel$ directions of momentum. The form given by Eq.~(\ref{cutoffDE}) is by no means mandatory and is motivated by convenience and the MF and LPA structures of the anisotropic propagator.
As we show below, the final numerical values obtained with the three cutoff function families are in fact similar and their differences are small when compared to the estimated error due to the truncation. 

\subsection{Anisotropic fixed points for $m=1$} \label{ssec:anisFPm1}
\begin{figure}
\includegraphics[width = \linewidth]{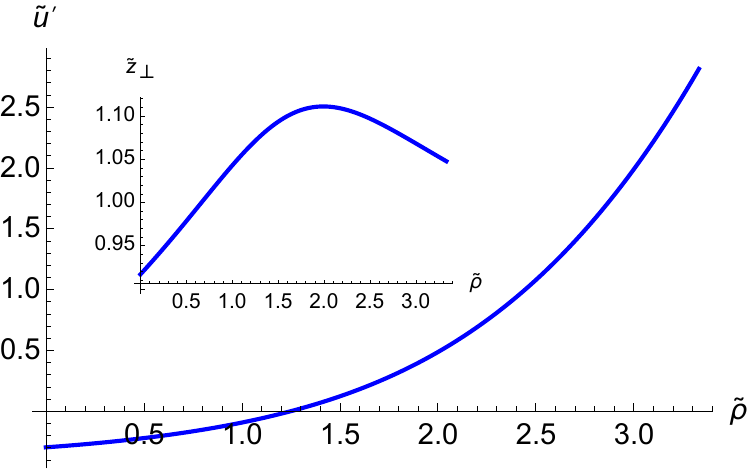}
\caption{The derivative of the fixed point effective potential $\tilde{u}'(\tilde{\rho})$ at the uniaxial ($m=1$) Lifshitz  point in $d=3$. The inset shows the corresponding shape of the function $\tilde{z}_\perp(\tilde{\rho})$. The plot corresponds to the exponential cutoff with ${\alpha=\alpha^{(PMS)}_{\eta_\perp}=0.95}$.} 
\label{u_FP}
\end{figure}

\begin{figure}
\includegraphics[width = \linewidth]{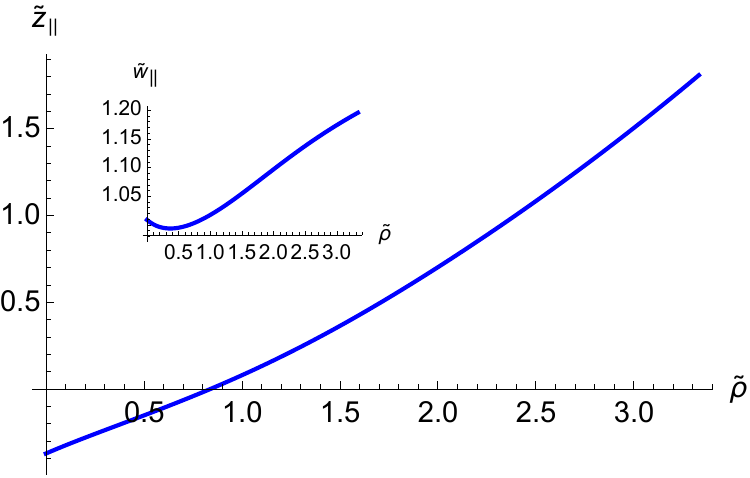}
\caption{The fixed point function $\tilde{z}_\parallel(\tilde{\rho})$ at the uniaxial ($m=1$) Lifshitz  point in $d=3$. The function is negative for small $\tilde{\rho}$. The inset shows the corresponding shape of the function $\tilde{w}_\parallel(\tilde{\rho})$. The plot corresponds to the exponential cutoff with ${\alpha=\alpha^{(PMS)}_{\eta_\perp}=0.95}$.}
\label{w_FP}
\end{figure}

With the aim of finding fixed-point solutions to Eq.~(\ref{uflow}) and Eq.~(\ref{floweq}) we discretize the $\tilde{\rho}$ variable (using a grid of  100 points) and map Eqs.~(\ref{uflow}), (\ref{floweq}) onto a system of ordinary integro-differential equations. Up to computing the integrals, the fixed point equation becomes then represented as a large set of algebraic equations, which can be studied with standard numerical tools relying on Newton's method and its generalizations. 
In practice, instead of Eq.~(\ref{uflow}) we use its derivative with respect to $\tilde{\rho}$. 
Details of the numerical procedure are explained in the Appendix, for a thorough discussion of its different aspects in the isotropic case, see \cite{Chlebicki_2024}. To identify the Lifshitz fixed point in the physically most relevant case $(d,m)=(3,1)$, we take advantage of the LPA correspondence described in Sec.~\ref{ssec:LPA}. We first consider dimensionality $d=3.5$, where the LPA approximation is relatively accurate, and where the anisotropic Lifshitz point is expected to be close to the standard isotropic $O(N)$ symmetric fixed point in dimensionality $d=3$. We use the latter and the value $\theta=1/2$ as the initial condition for finding the $d=3.5$ Lifshitz point. We subsequently gradually reduce $d$ using the previously computed fixed point as the new initial condition. This stepping is fully analogous to the procedure implemented for the $O(N)$ models e.g. in Refs.~\cite{Chlebicki_2021, Chlebicki_2022}. This procedure works because in this approach, the dimensionality enters as a parameter and does not appear manifestly in vectors dimensionality or as a discretization of space. The fixed-point profiles obtained this way at $d=3$ are plotted in Figs.~\ref{u_FP} and \ref{w_FP}. 
The functions $\tilde{u}'(\tilde{\rho})$ and $\tilde{z}_\perp(\tilde{\rho})$ are qualitatively similar to their counterparts studied previously for the isotropic $O(N)$ classes.  Interestingly, the function $\tilde{z}_\parallel(\tilde{\rho})$ is negative only for small values of $\tilde{\rho}$.   

\begin{figure}
\includegraphics[width = \linewidth]{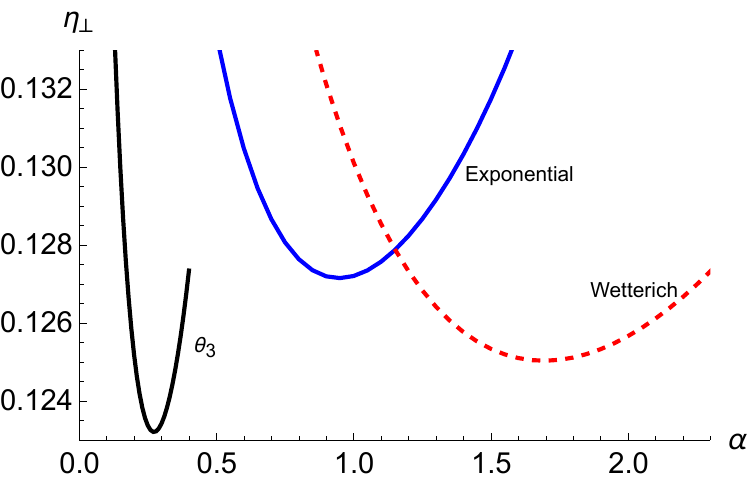}
\caption{Dependence of the exponent $\eta_\perp$ on the cutoff parameter $\alpha$ in the DE(2)+W$_\parallel(\rho)$ calculation, plotted for the three considered families of regulators. The PMS value of $\alpha$ as well as the shape of the curve differs substantially depending on the regulator type. However, the PMS value of the exponent differs by the relatively small amount of around 3\% depending on the regulator. } 
\label{eta_vs_alpha}
\end{figure}

\begin{figure}
\includegraphics[width = \linewidth]{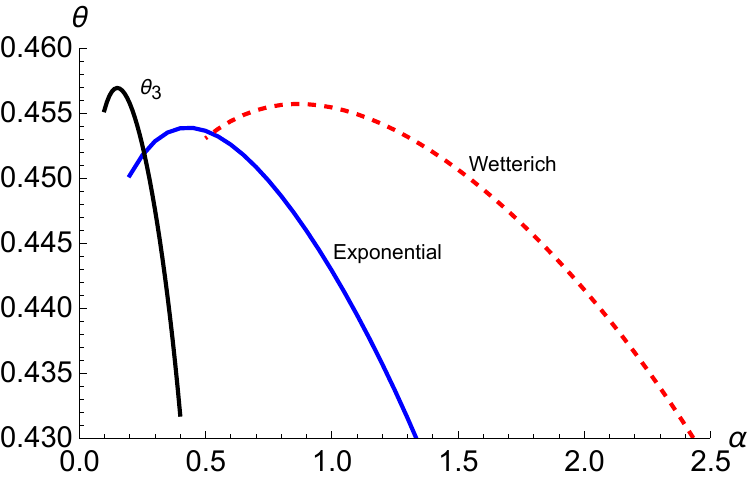}
\caption{Dependence of the anisotropy exponent $\theta$ on the cutoff parameter $\alpha$ in the DE(2)+W$_\parallel(\rho)$ calculation, plotted for the three considered families of regulators. The PMS value of $\alpha$ as well as the shape of the curve differs substantially depending on the regulator type; the PMS value of the exponent is however almost insensitive with respect to the regulator choice. } 
\label{theta_vs_alpha}
\end{figure}

Having identified the uniaxial Lifshitz fixed point in $d=3$, for each of the considered families of regulators, we scan the dependence of the exponents $\eta_\perp$ and $\theta$ on the cutoff parameter $\alpha$. Such dependencies  would not occur in an exact calculation and appear due to truncation. We implement the principle of minimal sensitivity (PMS) \cite{Canet_2003} to choose the values of $\alpha$, where the observables in question are (locally) least sensitive to the choice of $\alpha$. For a discussion of the methodology of PMS, its relation to minimal violation of the conformal Ward identity, convergence of the derivative expansion and the nature of its small parameter, see Refs.~\cite{Balog_2019,Balog_2020, Polsi_2022_2,Delamotte2024,Cabrera2024}. We present the plots of $\eta_\perp(\alpha)$ and $\theta(\alpha)$ in Figs.~\ref{eta_vs_alpha} and \ref{theta_vs_alpha}.
For the final values in $(d,m)=(3,1)$ we obtain: 
\be 
\theta \approx  0.455\;, \;\;\;\; \eta_\perp \approx 0.125\;, \;\;\; 
\eta_\parallel  \approx -0.12 \;,
\ee 
where, for $\theta$ and $\eta_\perp$ we averaged over the PMS values obtained with the three families of regulator. 
Our value of $\eta_\parallel$ follows from the scaling relation \cite{Diehl_2002} 
\be
\theta =\frac{2-\eta_\perp}{4-\eta_\parallel}\;
\ee 
already implicit in Eq.~(\ref{scaling}). For further discussion of the errors and comparison to values obtained from other approaches, see Sec.~\ref{ssec:ErEs} and Table \ref{tab:mainRes}. 

We note the negative sign of $\eta_{\parallel}$, which agrees with predictions of the $\epsilon$-expansion \cite{Diehl_2000} (at order $\epsilon^2$), but not the $1/N$ expansion \cite{Shpot_2001} - compare Table \ref{tab:mainRes}.  We emphasize that accuracy of all these estimates is by no means comparable to the precision reached for the conventional exponents pertinent to the isotropic fixed points, which is reflected by the scatter of the numbers collected in Table \ref{tab:mainRes} as well as the magnitude of error bars. 

As concerns the nonperturbative RG methodology, we believe that the present accuracy of the derivative expansion truncation should be compared to the DE(2) level of the conventional isotropic case. Reaching the complete DE(4) precision is beyond the scope of the present work. We note that the uncertainty estimate of the anomalous dimensions along the lines developed in \cite{Balog_2019, Polsi_2020, Balog_2020, Polsi_2022_2} without elevating the calculation to the complete DE(4) truncation is only possible by using the LPA value  $\eta_\perp=\eta_\parallel =0$. The situation is somewhat different for the correlation length exponents discussed below. We postpone the discussion of errors to Sec.~\ref{ssec:ErEs}.   

\subsection{Correlation lengths and crossover exponents} \label{ssec:CorLenCrossExp}
The values of $\theta$ and $\eta_\perp$ follow directly from the fixed-point solution, as described in the preceding section. Further critical exponents can be obtained from the leading eigenvalues of the RG transformation linearized around the fixed point. To construct the latter we introduce the ``vector" of parametrizing functions: 
\be 
\mathcal{F}(\tilde{\rho}):=\{  \tilde{u}(\tilde{\rho}),\tilde{z}_\perp(\tilde{\rho}), \tilde{z}_\parallel(\tilde{\rho}),\tilde{w}_\parallel(\tilde{\rho}) \}
\ee
and define the stability matrix 

\be
\mathcal{M}_{(i,\tilde{\rho}_1), (j,\tilde{\rho}_2)}:=\frac{\partial\big(\partial_t \mathcal{F}_i(\tilde{\rho}_1)\big)}{\partial \mathcal{F}_j(\tilde{\rho}_2)}\bigg|_{\eta_\perp^*,\theta^*,\mathcal{F}^*\;,}
\ee
where $\mathcal{F}^*$ is the vector of fixed-point functions. This represents the variant of the flow equations linearized around $\mathcal{F}^*$. After discretizing the field $\tilde{\rho}$, we extract the eigenvalues of thus obtained stability matrix.

\begin{figure}
\includegraphics[width = \linewidth]{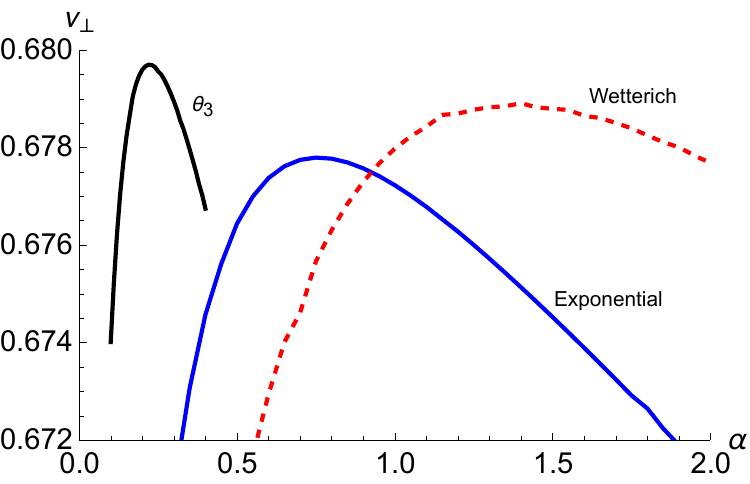}
\caption{Dependence of the $\nu_\perp$ exponent on the cutoff parameter $\alpha$ in the DE(2)+W$_\parallel(\rho)$ calculation, plotted for the three considered families of regulators. The PMS value of $\alpha$ as well as the shape of the curve differs substantially depending on the regulator type; the PMS value of the exponent is however almost insensitive with respect to the regulator choice, the difference being of order $0.3\%$ of the value of $\nu_{\perp}$. } 
\label{nu_vs_alpha}
\end{figure}

\begin{figure}
\includegraphics[width = \linewidth]{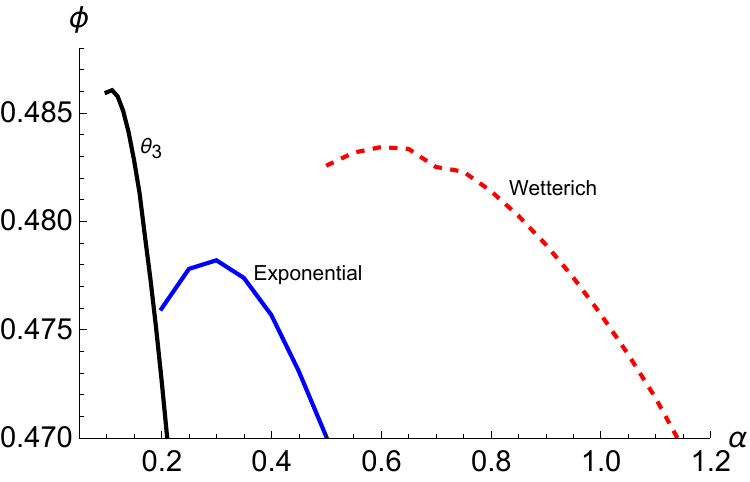}
\caption{Dependence of the crossover exponent $\phi$ on the cutoff parameter $\alpha$ in the DE(2)+W$_\parallel(\rho)$ calculation, plotted for the three considered families of regulators. The PMS value of $\alpha$ as well as the shape of the curve differs substantially depending on the regulator type; the PMS value of the exponent differs by less than $2\%$ of the value of $\phi$. } 
\label{phi_vs_alpha}
\end{figure}

In accordance with expectations, our calculation always yields two negative eigenvalues (denoted hereafter as $e_1$ and $e_2$) corresponding to the two relevant eigenperturbations. We identify the $\nu_\perp$ exponent as $\nu_\perp=e_1^{-1}$ and the crossover exponent $\phi$ \cite{Cardy_1996, Diehl_2002} as $\phi=e_2/e_1$. We perform the scan of $\alpha$ dependencies of these two quantities in a procedure analogous to the one implemented for $\eta_\perp$ and $\theta$ in Sec.\ref{ssec:anisFPm1} - see Figs.~\ref{nu_vs_alpha} and \ref{phi_vs_alpha}. 
For the PMS values averaged over the three implemented cutoffs, we find: 
\begin{equation}
\nu_\perp\approx 0.679\;, \;\;\;\; \phi\approx 0.48\;, \;\;\;\; \nu_\parallel \approx 0.3091 \;, 
\end{equation} 
where the value of $\nu_\parallel;$ follows from $\nu_\parallel =\theta \nu_\perp$ using PMS values of $\theta$ and $\nu_\perp$.

\begin{widetext}

\begin{table}[h]
\begin{tabular}{|c|c|c|c|c|c|c|}
\hline
                                                                       & $\theta$        & $\eta_\perp$    & $\nu_\perp$     & $\eta_\parallel$ & $\nu_\parallel$    & $\phi$        \\ \hline
MF                                                                     & 1/2             & 0               & 1/2             & 0                & 1/4                & 1/2             \\ \hline
$\mathcal{O}(\epsilon)$ \cite{Diehl_2000}            & 1/2             & 0               & 0.625           & 0                & 0.313              & 0.625         \\ \hline
$\mathcal{O}(\epsilon^2)$ \cite{Diehl_2000}          & 0.47            & 0.039           & 0.709           & -0.019           & 0.348              & 0.677         \\ \hline
Improved $\mathcal{O}(\epsilon^2)$ \cite{Shpot_2001} & 0.47            & 0.124           & 0.746           & -0.019           & 0.348              & 0.715         \\ \hline
1/N \cite{Shpot_2005, Shpot_2012}                   & 0.451           & 0.306           & 0.726           & 0.223            & 0.266              & -             \\ \hline
LPA (Ref.~\cite{Zdybel_2021})                   & 1/2             & 0               & 0.63            & 0                & -                  & -             \\ \hline
\bf{DE(2)+W}\bf{$_\parallel(\rho)$}                                              & \bf{0.455$\pm$0.011} & \bf{0.125$\pm$0.031} & \bf{0.679$\pm$0.012} & \bf{-0.12$\pm$0.03}   & \bf{0.3091$\pm$ 0.0093} & \bf{0.48$\pm$0.12} \\ \hline
\end{tabular} 
\caption{Comparison of the critical exponents controlling the uniaxial Lifshitz point in $d=3$ obtained from the DE(2)+W$_\parallel (\rho)$ truncation and from  other theoretical approaches - the $\epsilon$-expansion around $d=4+\frac{1}{2}$ and the $1/N$ expansion, see the main text in Sec.~\ref{ssec:CorLenCrossExp} for discussion.} \label{tab:mainRes}
\end{table} 
\end{widetext}

Our final estimates for the critical exponents at the uniaxial Lifshitz point in $d=3$ are collected and compared to those obtained from other approaches in Table \ref{tab:mainRes}. The agreement may be discussed only at a qualitative level. As a general observation, we note that for most cases our numbers compare relatively  favorably with those of the $\epsilon$-expansion and are often completely off to those from the $1/N$ expansion - see Table \ref{tab:mainRes}. As concerns the value of $\theta$, there is agreement between all the discussed  approaches indicating estimates in the range 0.45-0.47, somewhat below the classical value 1/2. For $\eta_\perp$ our result is (presumably by coincidence) very close to one of the predictions from the $\epsilon$-expansion and differs from the one from $1/N$ by a factor larger than 2. The values of $\nu_\perp$ predicted by the three approaches compare well. We note that our uncertainty for this exponent reaches below 2\% and is significantly lower than for some of the other critical indices (see also Sec.~\ref{ssec:ErEs}). For $\eta_\parallel$ we obtain a negative value way higher than the one from the $\epsilon$ expansion (but of the same sign). All the earlier results for $\eta_\parallel$ are way outside our error bars and there is no sign of agreement between any of the discussed predictions. In contrast, for $\nu_\parallel$ and $\phi$ the quoted numbers are in a qualitatively comparable range. 

\subsection{Error estimates}\label{ssec:ErEs}
Our error bar estimate methodology is based  on the insights of Refs.~\cite{Balog_2019, Balog_2020, Polsi_2020, Polsi_2021}, which identify a small parameter of the DE and propose, as well as test, a procedure relying on comparison of results obtained at  two consecutive orders of the systematic DE. Within this framework, the error of a quantity $\kappa$ obtained at order $\partial^n$ of the DE is conservatively estimated as 
\begin{equation}
\Delta \kappa^{(n)}=\frac{1}{4}|\kappa^{(n)}-\kappa^{(n-2)}|\;. 
\end{equation} 
As we previously emphasized, the analysis of Sec.~\ref{ssec:anisFPm1} and \ref{ssec:CorLenCrossExp}  should be understood as remaining at the level of DE(2), such that $n=2$ and $\kappa^{(n-2)}$ corresponds to the LPA prediction, where the anomalous dimensions are absent. In the present situation, this leads to rather large (25\%) relative error bars for  $\eta_\perp$ and $\eta_\parallel$. The same concerns the crossover exponent $\phi$, since LPA suppresses one of the relevant eigendirections related to the gradient term $Z_\perp$, and we adopt $e_2^{(LPA)}=0$ as the second  relevant RG eigenvalue at LPA level. This situation stands in contrast to the other exponents ($\nu_\perp$ in particular), where the relative errors are significantly smaller and reach the accuracy of around 2.4\% for $\theta$, 1.8\% for $\nu_\perp$ and 3\% for $\nu_\parallel$ - see Table \ref{tab:mainRes}. We also note that the estimated uncertainties from truncation are higher as compared to the one anticipated from the scatter of the PMS values obtained with the three regulator families - See Sec.~\ref{ssec:anisFPm1} and \ref{ssec:CorLenCrossExp}. 

\section{Summary and outlook}\label{sec:concl}
In recent years the nonperturbative RG methodology has been developed to become a computational tool capable of delivering {\it inter alia} high-precision numerical estimates of universal critical quantities. It has been extensively tested for the isotropic $O(N)$ models, where predictions from other methods are amply available. In the present study, we adapted and implemented the approach to a far more demanding situation involving anisotropic scale invariance, where no results from high-order perturbation theory or conformal theory are available, and where the expansion around the upper critical dimension appears more questionable. Working directly in $d=3$, we identified the functional RG fixed point controlling the uniaxial Lifshitz critical behavior and extracted a related set of critical exponents. We emphasize that a highly precise evaluation of these, in particular controlling the error bars, remains an open (and rather long-standing) 
problem, which in our opinion is partially resolved in the present study only for $\theta$, $\nu_\perp$, and $\nu_\parallel$. Our results unequivocally confirm the existence of the non-classical Lifshitz point in $d=3$ with the anisotropy index $\theta\approx 0.46$; a value not far from the predictions from the $\epsilon$-expansion around the upper critical dimension $d=4+\frac{1}{2}$ ($\theta\approx 0.47$), and indicate that the deviation from the MF prediction $\theta=1/2$ is actually not dramatic. Our prediction for the correlation length exponents (in particular for $\nu_\perp$) also compares reasonably with those of the $\epsilon$-expansion (notably the agreement is better with the "raw" variant of the $\epsilon$ expansion). We attribute this to a low level of credibility of the $\epsilon$-expansion, which is performed around $d=4+\frac{1}{2}$ and pushed only to order $\epsilon^2$. In fact, we consider our estimates to be more reliable than the previous predictions, in particular, with regards to those cases where we are able to provide relatively narrow error bars ($\theta, \nu_\perp, \nu_\parallel$).
Concerning the anomalous dimensions, we find qualitative agreement of $\eta_\perp$ with the $\epsilon$-expansion (in this case  the agreement is better with the resummed version of the $\epsilon$ expansion). For $\eta_\parallel$ we disagree with the $1/N$-expansion on both the sign and order of magnitude; while we agree with the $\epsilon$ expansion on the sign, but differ on the value by a factor of around 6. Concerning the signs, we also agree with the earlier, simplified and less controlled studies from nonperturbative RG.     

In our opinion going beyond the present truncation level and reaching the precision of the DE(4) approximation is a demanding, but  viable enterprise, which would deliver much narrower error estimates of all the critical exponents in question. The same is true concerning  an extension to arbitrary values of $d$ and $m$, except at the vicinity of the lower critical dimension. We are somewhat more reserved about generalization to arbitrary values of $N$, where going beyond the DE(2)+W$_\parallel (\rho)$ truncation in a controlled fashion represents a formidable challenge.   

\begin{acknowledgments}
  P.J. acknowledges support from the Polish National Science Center via grant 2021/43/B/ST3/01223 and thanks Ivan Balog for discussions at the initial stage of the project. G.D.P. acknowledges support from the Programa de Desarrollo de las Ciencias Básicas (PEDECIBA) and from the grant of number FCE-3-2024-1-180709 of the Agencia Nacional de Investigación e Innovación (Uruguay). We are grateful to Nicolas Wschebor for useful remarks on the initial version of the manuscript as well as numerous discussions on related topics.
\end{acknowledgments} 

\appendix

\section{Numerical method}\label{Ap:Numeric}

We describe in this section the details of the numerical implementation used to find the fixed point and compute the universal critical quantities. The starting point are the flow equations  for the different ansatz functions obtained by means of the DE procedure and given in the supplemental material. To treat these expressions numerically there are two key steps: i) Evaluating the flow expressions which involve partial derivatives in the field and internal momenta integration in two independent variables and ii) finding the fixed point solution from a starting initial guess and computing critical quantities of interest from the fixed point solution.

\subsection{Evaluating flow expressions}

The expressions to be evaluated involve partial derivatives of the functions in the field. To proceed we consider a uniform grid in the variable $\rho$ with $N_\rho=101$ points and with a step $\delta\rho=1/30$. The derivatives are then discretized in the grid with $2m+1$ points centered at each point, except at the $m$ points closest to the each edge where the discretized derivatives are taken as centered as possible.

The flow equations for the ansatz functions involve up to two derivatives which implies that at least three-point derivatives are needed. In the final implementation we used five-point derivatives since results are already stable and considering seven or nine-points derivatives does not alter the results presented here.

Another key aspect of the numerical procedure is the evaluation of the momenta integrals such as the one appearing in Eq.\eqref{uflow} or the ones in the supplemental material. A generic integral $I(\rho)$ appearing in the equations is of the form:
\begin{equation}
    I(\rho)=\mathcal{V}_{d,m}\int_0^\infty\int_0^\infty d\tilde{q}_\perp d\tilde{q}_\parallel \tilde{q}_\perp^{d-m-1}\tilde{q}_\parallel^{m-1}f(\tilde{q}_\perp^2,\tilde{q}_\parallel^2,\rho),
\end{equation}
where $f$ is a function suppressed at large $\tilde{q}_\perp$ or $\tilde{q}_\parallel$ due to the regulator and $\mathcal{V}_{d,m}$ is set to $1$ since it can be absorbed in a redefinition of the field and the ansatz functions. The integrals are performed using the Cuhre algorithm of degree 13 from the Cuba library \cite{Hahn2005}. The upper bound of the integrals were set as $\tilde{q}_\perp^{max}=3.2$ and $\tilde{q}_\parallel^{max}=1.8$. Absolute and relative error were chosen to $\varepsilon^{abs}=\varepsilon^{rel}=10^{-4}$ which yielded same results as higher precision but with a better performance time-wise. Since the Wetterich regulator is ill-defined numerically at 0 momenta, a Taylor expansion up to $o(\tilde{q}_\perp^{P_\perp})$ and $o(\tilde{q}_\parallel^{P_\parallel})$ was considered for momentum $\tilde{q}_\perp\leq Q^{th}_{\perp}$ and $\tilde{q}_\parallel\leq Q^{th}_{\parallel}$. We fixed $Q^{th}_{\perp} =Q^{th}_{\parallel}=0.2$ just to be far from numerical instability while having stable integrals. Additionally, the power considered for the Taylor expansion was set to $P_\perp=P_\parallel=8$ although numerical convergence was obtained for smaller powers of the momenta.

\subsection{Solving for the fixed point and computing critical exponents}

As mentioned in the text, to find the fixed point for the Lifshitz model we used as an initial condition the fixed point for the isotropic case in reduced dimensionality. We then proceed to slowly reduce the dimension of the Lifshitz system from dimension $d=3.5$ to $d=3$. Subsequently, we proceeded to vary the overall scale of the regulator $\alpha$.

In order to find a particular fixed point from an initial condition, we performed a straightforward Newton-Raphson (NR) method computing the stability matrix numerically and imposing that $\beta$-functions or flows should be negligible enough in order to guarantee for stable results. This, in general, was implemented by imposing a measure of distance and iterating until the norm of the $\beta$-function vector $\vec{\beta}\equiv(\beta_{U'_0},\beta_{U'_1},\dots,\beta_{W_{a\parallel,N_\rho-1}})$ was smaller than a threshold $\epsilon^{dist}$.

The fixed-point functions $Z_\perp$ and $W_{a\parallel}$ were normalized as $Z_\perp|_{\rho_{norm}}=W_{a\parallel}|_{\rho_{norm}}=1$. The normalization point used was $\rho_{norm}=20\,\delta\rho$. This normalization scheme was implemented throughout the NR procedure which allowed to compute $\eta_\perp$ and $\theta$ as a function of the ansatz functions. Once the fixed point was attained, the fixed point critical exponents $\eta_\perp^*$ and $\theta^*$ were obtained (see Figs.\,\ref{eta_vs_alpha}-\ref{theta_vs_alpha}).

Once the fixed point is obtained by means of the NR procedure, we perform a linear stability analysis which uses the stability matrix $\mathcal{M}$ already computed in the NR procedure. We then compute the eigenvalues of matrix $\mathcal{M}$ of size $(4N_\rho-2)\times(4N_\rho-2)$. There are only two relevant eigenvalues $e_1$ and $e_2$ which relate to the correlation length exponent $\nu_\perp$ and to the crossover exponent $\phi$ as $\nu_\perp=-e_1^{-1}$ and $\phi=e_2/e_1$ (see Figs.\,\ref{nu_vs_alpha}-\ref{phi_vs_alpha}).

In order to find a particular fixed point from an initial condition, we performed a straightforward Newton-Raphson (NR) method computing the stability matrix numerically and imposing that $\beta$-functions or flows should be negligible enough in order to guarantee for stable results. This, in general, was implemented by imposing a measure of distance and iterating until the norm of the $\beta$-function vector $\vec{\beta}\equiv(\beta_{U'_0},\beta_{U'_1},\dots,\beta_{W_{a\parallel,N_\rho-1}})$ was smaller than a threshold $\epsilon^{dist}$.

We highlight, once again, that all our results were explored by changing parameters ($N_\rho$, $\delta\rho$, $m$, $\tilde{q}_\perp^{max}$, $\tilde{q}_\parallel^{max}$, $\varepsilon^{abs}$, $\varepsilon^{rel}$, $\epsilon^{dist}$, $P_\perp$, $P_\parallel$, $Q^{th}_{\perp}$ and $Q^{th}_{\parallel}$) to increase performance and guarantee stability of results.

\bibliography{bibliography}

\end{document}